# Application of the EXtrapolated Efficiency Method (EXEM) to infer the gamma-cascade detection efficiency in the actinide region

Q. Ducasse[1,2], B. Jurado[1,*], L. Mathieu[1], P. Marini[1], B. Morillon[3], M. Aiche[1], I. Tsekhanovich[1]

1) CENBG, CNRS/IN2P3-Université de Bordeaux, Chemin du Solarium B.P. 120, 33175 Gradignan, France
2) CEA-Cadarache, DEN/DER/SPRC/LEPh, 13108 Saint Paul lez Durance, France
5) CEA DAM DIF, 91297 Arpajon, France

**Abstract:**
The study of transfer-induced gamma-decay probabilities is very useful for understanding the surrogate-reaction method and, more generally, for constraining statistical-model calculations. One of the main difficulties in the measurement of gamma-decay probabilities is the determination of the gamma-cascade detection efficiency. In [Nucl. Instrum. Meth. A 700, 59 (2013)] we developed the Extrapolated Efficiency Method (EXEM), a new method to measure this quantity. In this work, we have applied, for the first time, the EXEM to infer the gamma-cascade detection efficiency in the actinide region. In particular, we have considered the $^{238}U(d,p)^{239}U$ and $^{238}U(^{3}He,d)^{239}Np$ reactions. We have performed Hauser-Feshbach calculations to interpret our results and to verify the hypothesis on which the EXEM is based. The determination of fission and gamma-decay probabilities of $^{239}Np$ below the neutron separation energy allowed us to validate the EXEM.

## 1. Introduction

Neutron-induced reaction cross sections of short-lived nuclei are important in several domains such as fundamental nuclear physics, nuclear astrophysics and applications in nuclear technology. However, very often these cross sections are extremely difficult (or even impossible) to measure due to the high radioactivity of the targets involved. The surrogate-reaction method was developed to overcome this difficulty. In this method, the same decaying nucleus as in the neutron-induced reaction of interest is formed via an alternative or surrogate reaction (typically a transfer or inelastic scattering reaction) involving a target nucleus that is easier to produce and handle than the target needed in the desired neutron-induced reaction. In the surrogate-reaction method, the desired neutron-induced cross section is "simulated" by multiplying the decay probability measured in the surrogate reaction with the calculated cross section for the formation of a compound nucleus after neutron absorption. For more details on the surrogate-reaction method see the review article [1]. Studies performed in the last decade have shown that fission cross sections obtained via the surrogate-reaction method are generally in good agreement with the corresponding neutron induced data, see e.g. [2] and other examples given in [1]. However, discrepancies as large as a factor 10 have been observed when comparing radiative-capture cross sections of rare-earth nuclei obtained in surrogate and neutron-induced reactions [3, 4]. We have shown in [5] that to understand these results it is necessary to measure simultaneously the fission and gamma-decay probabilities induced by different surrogate reactions, and that these measurements represent a strong constrain for the statistical model. In [5], we measured for the first time simultaneously the two decay probabilities for the $^{238}$U(d,p) surrogate reaction. In a later experiment, we measured simultaneously fission and gamma decay probabilities induced by $^{3}$He + $^{238}$U transfer reactions.

---
*  jurado@cenbg.in2p3.fr

The technique to determine transfer-induced fission probabilities is rather well mastered nowadays. A thorough description of the methodology and the associated uncertainties can be found in [6]. However, the situation is more complicated for the determination of transfer-induced gamma-decay probabilities, in particular when one is interested in an excitation-energy domain where the gamma-decay and fission channels compete. The gamma-decay probability measures the probability that an excited nucleus $A^*$ decays through a gamma-ray cascade, i.e. that nucleus $A^*$ de-excites by emitting gamma rays only. Above the neutron separation energy $S_n$ this probability decreases very rapidly with excitation energy $E^*$, because of the competition with the neutron-emission channel. Therefore, the gamma-decay probability is typically measured up to $E^* \approx S_n + 1$ MeV. The gamma-decay probability $P_\gamma$ as a function of $E^*$ of nucleus $A^*$ produced by a transfer reaction $X(y,w)A^*$ can be obtained as:

$$P_\gamma(E^*) = \frac{N_\gamma^C(E^*)}{N^S(E^*) \cdot \varepsilon_\gamma(E^*)} \quad (1)$$

where $N^S$ is the total number of detected ejectiles $w$, i.e. the number of formed excited nuclei $A^*$. $N_\gamma^C$ is the number of ejectiles $w$ detected in coincidence with a gamma cascade and $\varepsilon_\gamma$ is the gamma-cascade detection efficiency. The quantity $N_\gamma^C / \varepsilon_\gamma$ gives the number of produced nuclei $A^*$ that decay via the emission of a gamma cascade. The two main difficulties in the determination of the gamma-decay probability are the subtraction of the background gamma rays coming from the fission fragments and the determination of the gamma-cascade detection efficiency. The procedure to subtract the gamma-ray background is shown in [5]. In this work we concentrate on the determination of $\varepsilon_\gamma$.

The gamma-cascade detection efficiency $\varepsilon_\gamma$ depends on the multiplicity $M\gamma$ and the energies $E\gamma$ of the gammas of the cascade. These two latter quantities depend in turn on $E^*$. Moreover, in the quasi-continuum and continuum regions, the cascade paths can be very different from one event to the other, even if $E^*$ is the same. This makes rather difficult the determination of the gamma-cascade detection efficiency at $E^*$ close to the neutron separation energy $S_n$, which is the region of interest. Note that this is not a specific problem of transfer-induced measurements, but it is also found in neutron-induced radiative-capture measurements. In these experiments the Pulse-Height Weighting-function Technique (PHWT) is used to determine the gamma-cascade detection efficiency. The principle of this technique is described for example in [7] and references therein. The PHWT is quite complicated as it requires to determine the response functions of the detector array for many incident gamma-ray energies $E_\gamma$, ranging from a few hundred keV to about $S_n + 1$ MeV in steps of a few tens of keV. For the correct application of the PHWT, it is also necessary to include the contribution from gamma rays with energies below the electronic threshold, which needs reproducing the nucleus de-excitation process with a model and simulating the detection setup. In addition, the PHWT requires a setup that minimizes the cross talk between the gamma detectors. In [8], we presented an alternative method for determining the gamma-cascade detection efficiency in surrogate-reaction experiments called the EXtrapolated-Efficiency Method, EXEM. The EXEM is much simpler than the PHWT, it does not require the knowledge of the detector response matrix or of the measured spectrum below the applied gamma-energy threshold and is not affected by cross-talk effects. In [8], we considered the rare-earth nucleus $^{176}$Lu*, for which the variation of the efficiency with excitation energy was rather weak. The dependence of the efficiency on excitation energy is expected to be different for heavier nuclei because of changes in the level densities and gamma-ray strength functions. Therefore, in this work we investigate the EXEM in the actinide region. In particular, we have considered the $^{239}$U* nucleus formed in the $^{238}$U(d,p) reaction and the $^{239}$Np* nucleus formed in the $^{238}$U($^3$He,d) reaction. The EXEM method can be applied to any type of gamma detector, but so far it has only been tested with an ensemble of $C_6D_6$ liquid scintillators in [8]. In this work, we apply the EXEM to an array of high-efficiency NaI scintillators.

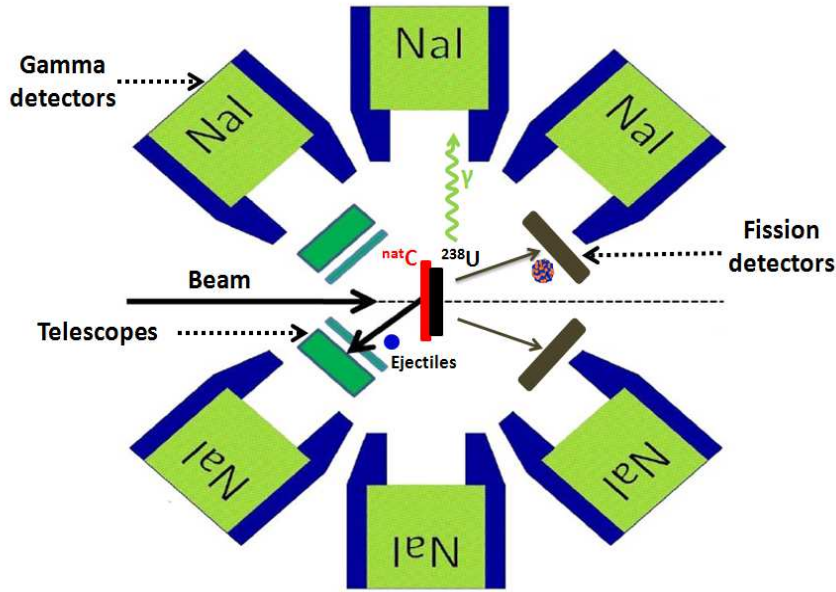

**Figure 1:** Schematic view of the setup used at the Oslo cyclotron for the simultaneous measurement of fission and gamma decay-probabilities.

2. **Experimental setup**

The measurements were performed at the Oslo Cyclotron Laboratory that provided a deuteron beam of 15 MeV energy with an intensity of about 4 enA, and a $^3$He beam of 24 MeV with an intensity of about 0.4 enA. We used a high-quality metallic $^{238}$U target, with 99.5% isotopic purity, produced by the GSI target laboratory. It had a thickness of 260 µg/cm$^2$ and was deposited on a 40 µg/cm$^2$ natural carbon layer. The experimental setup is sketched in Fig. 1.

The multi-strip ΔE/E silicon telescope SiRi [9], covered with a 21 µm thick aluminum foil to stop fission fragments, was used to identify the ejectiles and measure their kinetic energies and angles. With this information it was possible to identify the decaying nucleus and determine its $E^*$. An ensemble of four PPACS [10], located at forward angles, was used to detect the fission fragments in coincidence with the ejectiles. This fission detector served to measure the fission probability and to remove the gamma-ray background coming from the de-excitation of the fission fragments. The reaction chamber housing SiRi and the $^{238}$U target was surrounded by the CACTUS array [11], constituted of 27 high-efficiency NaI detectors. CACTUS was used to detect gamma rays with energies ranging from few hundreds of keV to about 10 MeV, emitted in coincidence with the ejectiles.

### 3. Application of the EXEM to the $^{238}$U(d,p)$^{239}$U* reaction

As discussed in the introduction, to obtain the gamma-decay probability for the $^{238}$U(d,p) reaction we need to determine the number of formed $^{239}$U* nuclei that decay through a gamma-ray cascade, $N_\gamma^C(E^*)$. The gamma-ray detection efficiency of the CACTUS array is about 14.5% at 1.33 MeV gamma-ray energy. Therefore, in most cases, we detected only one gamma ray per cascade. For the few cases where more than one NaI detector was hit in one event, we randomly selected one detector signal amplitude in the offline data analysis. In that way, we avoided counting more than one gamma ray per cascade.

Contrary to radiative neutron-capture reactions, in a transfer reaction it is possible to populate states with excitation energies below the neutron separation energy. The neutron-rich nucleus $^{239}$U* neither fissions nor emits protons at $E^*$ below $S_n$. Therefore, the only possible de-excitation mode in this $E^*$ region is gamma decay. This implies that its gamma-decay probability is equal to 1:

$$P_\gamma(E^*) = 1 = \frac{N_\gamma^C(E^*)}{N^S(E^*) \cdot \varepsilon_\gamma(E^*)} \qquad \text{for } E^* < S_n \qquad (2)$$

From eq. (2) it follows that:

$$\varepsilon_\gamma(E^*) = \frac{N_\gamma^C(E^*)}{N^S(E^*)} \qquad \text{for } E^* < S_n \qquad (3)$$

Thus, for excitation energies below $S_n$, the gamma-cascade detection efficiency, $\varepsilon_\gamma(E^*)$, can be directly obtained from the ratio between the number of gamma-ejectile coincidences and the total number of detected ejectiles. The ratios measured when applying two different gamma-ray energy thresholds, 400 keV corresponding to the electronic threshold, and 1.5 MeV, are shown in Fig. 2. Obviously, the efficiency considerably decreases as the gamma-energy threshold increases. We see that in both cases the efficiency increases steeply with $E^*$ below $S_n$. Above $S_n$ the ratio decreases very rapidly because of the onset of neutron emission, which competes with gamma emission and leads to a strong reduction of $N_\gamma^C$. For the 400 keV threshold we observe an increase of the ratio above about $E^* = 5.2$ MeV. This is due to the contribution of gamma rays from $^{238}$U*, which is produced after neutron emission from $^{239}$U*. The gamma rays emitted by the residue nucleus $^{238}$U* have a maximum energy $E\gamma = E^* - S_n$. For $^{239}$U, $S_n = 4.8$ MeV and if we limit the measurement of the gamma-decay probability to a maximum excitation energy $E^* = 6.3$ MeV, the maximum energy of the gammas emitted by the residue $^{238}$U is $E\gamma = 1.5$ MeV. Figure 2 shows that the events coming from the $^{238}$U(d,pn) reaction are removed when the 1.5 MeV threshold is applied.

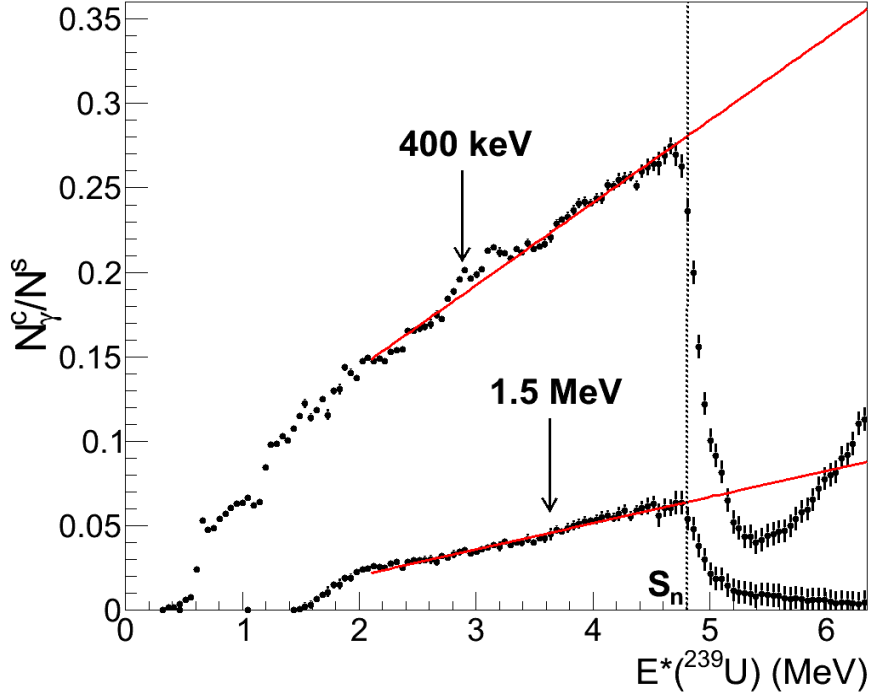

**Figure 2:** Ratio between the number of gamma-ejectile coincidences and the total number of detected ejectiles as a function of the excitation energy of $^{239}$U for two gamma-ray energy thresholds of 400 keV and 1.5 MeV. The ejectiles were detected at 126 degrees. The vertical dotted line indicates the neutron separation energy of $^{239}$U and the red solid lines are linear fits to the data in the $E^*$ interval [2 MeV; $S_n$].

For medium-mass and actinide nuclei in the region of continuum level densities there is no reason to expect a drastic change at $S_n$ of the characteristics of the gamma cascades (multiplicity and average gamma energy), and thus of $\varepsilon_\gamma(E^*)$. This is the main hypothesis on which the EXEM is based. The EXEM assumes that the dependence of the gamma-cascade detection efficiency $\varepsilon_\gamma$ on $E^*$ measured below $S_n$ can be extrapolated to excitation energies above $S_n$. Therefore, to determine the gamma-cascade detection efficiency we applied a linear fit to the ratio. The values of the fit function evaluated at $E^*$ above $S_n$ give the gamma-cascade efficiency $\varepsilon_\gamma$ needed to determine the gamma-decay probability. The uncertainty on $\varepsilon_\gamma$ above $S_n$ is given by the uncertainties on the fit parameters.

In Fig. 3 we show the values for the average gamma-ray energy $<E_\gamma>$ and gamma multiplicity $<M_\gamma>$ as a function of excitation energy. The experimental values for $<E_\gamma>$ have been obtained from the data after unfolding the measured spectra into the full-energy gamma-ray spectra according to the method of ref. [12]. The unfolded spectra at different excitation energies can be seen in the two-dimensional plot in Fig. 4. <$M_\gamma$> was indirectly deduced from the data via equation:

$$E^* = <E_\gamma> \cdot <M_\gamma> \qquad (4)$$

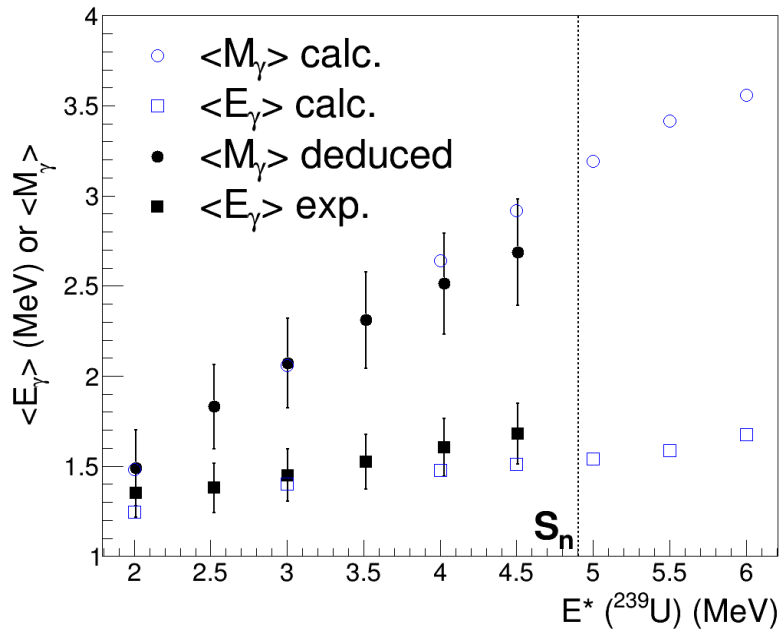

**Figure 3:** Average gamma-ray energy and multiplicity as a function of excitation energy for the gamma decay of $^{239}$U*. The full symbols have been obtained from the experimental data and the empty symbols are the results of the EVITA code. In both cases a gamma-ray energy threshold of 400 keV was used. The vertical dashed line represents the neutron separation energy $S_n$ of $^{239}$U.

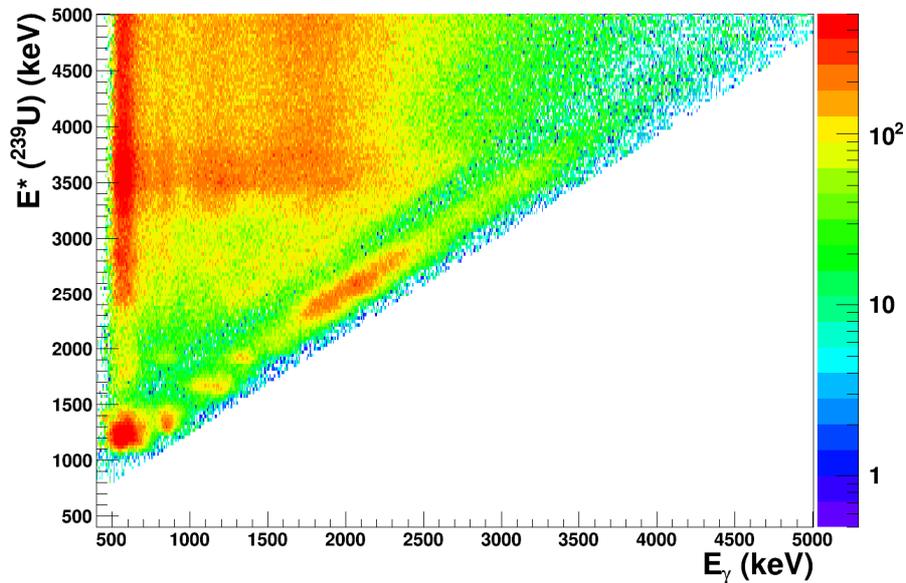

**Figure 4:** Coincidence matrix for the $^{238}$U(d,p) reaction representing the excitation energy of $^{239}$U versus the gamma-ray energy. The gamma-ray energies shown have been obtained by unfolding the measured gamma-ray spectra with the method of [12].

Fig. 3 shows that the gamma multiplicity increases much more strongly with $E^*$ than the average gamma energy. By comparing the curve of figure 2 with the 400 keV threshold and the deduced multiplicity of figure 3, it can be seen that the efficiency and the multiplicity increase at a similar rate of about 30% per MeV. This strong link between the efficiency and the multiplicity is expected theoretically. For a detector such as CACTUS with a relatively low gamma-ray detection efficiency $\alpha$, which depends rather weakly on the gamma-ray energy well above the threshold [12], we have:

$$\varepsilon_\gamma \approx \sum_{i=1}^{M_\gamma} \alpha_i(E_\gamma) \approx \alpha \cdot M_\gamma \qquad (5)$$

where $\alpha_i$ is the efficiency for detecting a gamma ray with a particular energy $E_\gamma$ from a cascade of multiplicity $M_\gamma$. Therefore, according to eq. (5), the $\varepsilon_\gamma$ of CACTUS is nearly proportional to $M_\gamma$ and the proportionality constant corresponds to the gamma detection efficiency $\alpha$. The data shown in figures 2 and 3 satisfy fairly well eq. (5). Indeed, we find a constant of proportionality between $\varepsilon_\gamma$ and $<M_\gamma>$ of about 0.1, which is rather close to the efficiency of CACTUS $\alpha \approx 0.13$ at $E_\gamma = 1.5$ MeV. We then conclude that the observed increase of the efficiency $\varepsilon_\gamma$ with excitation energy shown in figure 2 is essentially due to the increase of the multiplicity of the gamma cascades.

The values of $<E_\gamma>$ and $<M_\gamma>$ obtained with the EVITA code [13] are also shown in figure 3. EVITA is a Hauser Feshbach Monte-Carlo code that uses the same ingredients as the TALYS code [14]. The parameters of the EVITA code have been carefully tuned to reproduce the experimental neutron-induced data of $^{238}$U. The results of the calculations for $<E_\gamma>$ and $<M_\gamma>$ depend very weakly on the angular momentum and parity of the decaying nucleus $^{239}$U*. Therefore, the results of EVITA are also valid when the $^{239}$U* nucleus is produced via the $^{238}$U(d,p) reaction, where the transferred angular momentum is expected to be larger than in the neutron-induced reaction. The calculated values of $<E_\gamma>$ and $<M_\gamma>$ were directly computed with the calculated cascades to which we applied a threshold of 400 keV. We can see that the experimentally deduced quantities and the calculations agree well within the error bars. This indicates that eq. (4) is still rather well fulfilled despite the energy threshold. The calculation shows that for both quantities there is nearly no change in the slope above 2 MeV. It is thus well justified to extrapolate the functional form of $\varepsilon_\gamma(E^*)$ measured below $S_n$ to $E^*$ above $S_n$, which is the essential hypothesis behind the EXEM. The evolution of the calculated average gamma multiplicity of $^{239}$U differs significantly from the one of $^{176}$Lu shown in Fig. 4 of [8]. For the latter case, statistical-model calculations showed a much softer increase of the average multiplicity with excitation energy, which explains the observed nearly constant behaviour of the efficiency.

### 4. Application of the EXEM to the $^{238}$U($^3$He,d)$^{239}$Np* reaction

The neutron-rich $^{239}$Np nucleus is of particular interest since it can be used to verify the validity of the EXEM. The fission barrier of $^{239}$Np is lower than its neutron separation energy. Therefore, for this nucleus two decay modes, fission and gamma-decay, are possible below $S_n$ and the sum of the corresponding decay probabilities must be equal to 1. These two decay probabilities were measured in an independent way with the setup described in section 2. For the details on the determination of the fission probability we refer to [5]. As for the gamma-decay probability, to determine the quantity $N_\gamma^C$ of eq. (1) we applied several corrections. The details of the analysis procedure can be found in [5], here we recall only the main aspects. To remove the background of gamma rays originating from the $^{238}$U($^3$He,dn) reaction we applied a gamma-energy threshold of 1 MeV. The resulting coincidence events $N_\gamma^{C,tot}$ had still to be corrected from the background originating from prompt gamma rays emitted by the fission fragments. The corrected coincidence spectrum $N_\gamma^C$ was obtained as:

$$N_\gamma^C(E^*) = N_\gamma^{C,tot}(E^*) - \frac{N_{\gamma,f}^C(E^*)}{\varepsilon_f(E^*)} \tag{6}$$

where $N_{\gamma,f}^C$ is the number of gamma cascades detected in coincidence with a deuteron and a fission fragment, and $\varepsilon_f$ is the fission detection efficiency. The solid angle of the fission detector was determined with the help of a $^{252}$Cf source of known activity. The final fission efficiency included also the effects of the angular anisotropy of the fission fragments. We observed a very weak dependence of $\varepsilon_f$ with $E^*$. The final fission efficiency was (48.0 ± 3,5)%. The gamma-cascade detection efficiency was determined by applying the EXEM, as shown in Fig. 5. In this case, the ratio of gamma-ejectile coincidence events over total number of ejectiles shows a step decrease below $S_n$, at about 5 MeV. This decrease is due to the onset of fission. To determine the gamma-cascade detection efficiency, a linear fit was applied below the fission threshold in the region from 2 to 5 MeV and was extrapolated up to 7.3 MeV, well above $S_n$.

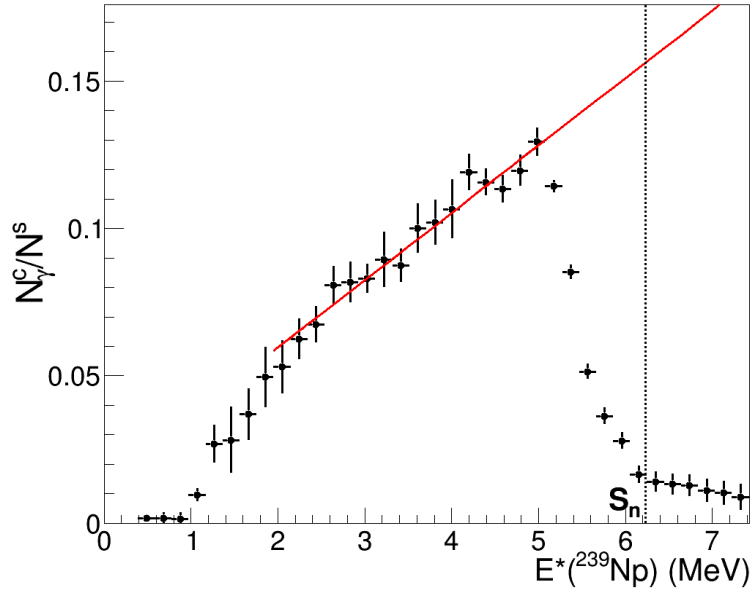

**Figure 5:** Ratio between the number of gamma-ejectile coincidences and the total number of detected ejectiles as a function of the excitation energy of $^{239}$Np for a gamma-ray energy threshold of 1 MeV. The vertical dotted line indicates the neutron separation energy $S_n$ of $^{239}$Np and the red solid line is a linear fit to the data in the $E^*$ interval [2; 5] MeV.

The results for the gamma-decay and fission probabilities are shown in Fig. 6. Both probabilities decrease at $S_n$ because of the competition with the neutron-emission channel. As expected, our fission probability is in good agreement with the neutron-induced fission probability given by ENDF/B-VII.1. This demonstrates that the fission probability and, consequently, the fission efficiency have been correctly determined. As shown in eq. (6), the fission efficiency is the key quantity to ensure the correct subtraction of the gamma-ray background coming from the fission fragments. Fig. 6 shows that the sum of the two decay probabilities is consistent with 1 below $S_n$. This result validates the analysis procedure for determining the fission and gamma-decay probabilities, and in particular the EXEM. Indeed, our results demonstrate that the linear increase of the efficiency measured between 2 and 5 MeV excitation energy can be extrapolated up to about 6.2 MeV, the $S_n$ of $^{239}$Np.

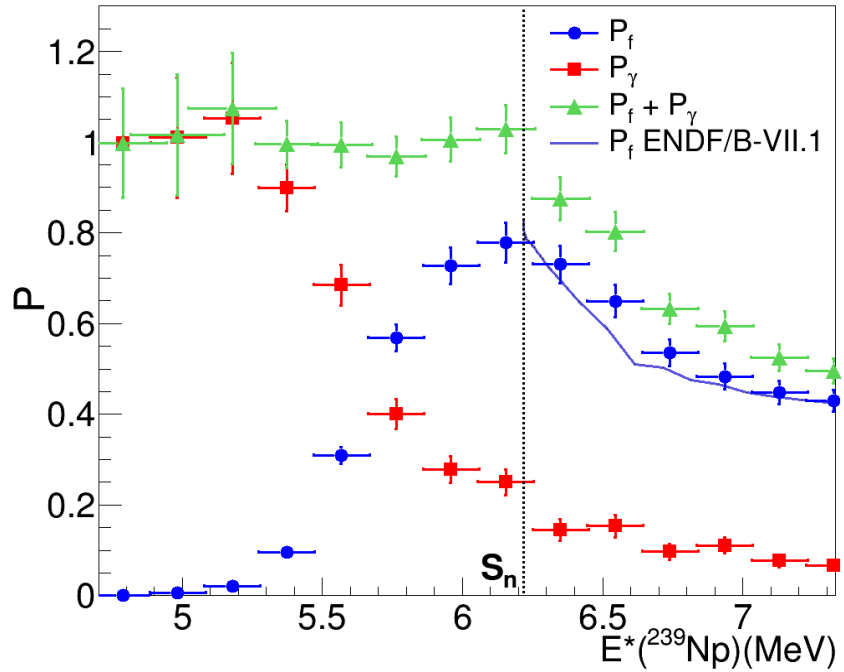

**Figure 6:** Fission (blue circles) and gamma-decay (red squares) probabilities as a function of the excitation energy of $^{239}$Np*. The sum of the two decay probabilities is represented by the green triangles. The full blue line is the neutron-induced fission probability given by the ENDF/B-VII.1 international library. The vertical dotted line indicates the neutron separation energy $S_n$ of $^{239}$Np.

5. **Conclusion**

The gamma-cascade detection efficiency is a key quantity for determining the gamma-decay probability. The Extrapolated Efficiency Method (EXEM), developed in [8], consists in extrapolating to higher excitation energies the excitation-energy dependence of the gamma-cascade detection efficiency measured below the neutron-emission threshold in a transfer-induced nuclear reaction. In this work, we have applied for the first time the EXEM to infer the gamma cascade detection efficiency in the actinide region. In particular, we have considered the $^{239}$U* and $^{239}$Np* nuclei produced in the $^{238}$U(d,p) and $^{238}$U($^{3}$He,d) reactions, respectively. In addition, here we have used high-efficiency NaI detectors instead of the $C_6D_6$ liquid scintillators used in [8]. Compared to the $^{176}$Lu rare-earth nucleus studied in [8], the actinide nuclei investigated in this work present a strong linear increase of the efficiency with excitation energy. Average multiplicities and gamma-ray energies extracted from the data and calculated with the Hauser-Feshbach code EVITA for the decay of $^{239}$U* show that this increase is mainly due to an increase of the gamma multiplicity of the cascades. The EVITA calculations also show that the functional form of the dependence on the excitation energy of the properties of the cascade, average multiplicity and gamma energy, remains essentially unchanged in an energy interval that ranges from several MeV below to about 1 MeV above the neutron separation energy. This justifies the essential hypothesis behind the EXEM. The application of the EXEM to the $^{239}$Np* nucleus is particularly interesting because for this nucleus the gamma-decay and fission channels are open below the neutron separation energy. We have found that the sum of the measured gamma decay and fission probabilities is very close to 1 below the neutron separation energy, demonstrating the validity of the EXEM. More precisely, our results demonstrate that the linear increase of the efficiency measured between 2 and 5 MeV excitation energy can be extrapolated up to about 6.2 MeV.

## Acknowledgments

We would like to thank the staff of the Oslo Cyclotron Laboratory for their support during the experiment and the GSI Target Laboratory for the production of the $^{238}$U target. We would also like to express our gratitude to Magne Guttormsen for providing us with the unfolded gamma-ray spectra. This work was supported by the European Commission within the 7$^{th}$ Framework Program through Fission-2010-ERINDA (Project No. 269499).